\documentclass[journal]{IEEEtran}

\usepackage[caption=true,font=normalsize,labelfont=sf,textfont=sf]{subfig}
\usepackage{pgfplots,filecontents}
\usepackage[cmex10]{amsmath}
\usepackage{amssymb}
\usepackage{algorithm,algorithmic}
\usepackage{array}
\usepackage{booktabs,comment}
\usepackage[caption=true,font=normalsize,labelfont=sf,textfont=sf]{subfig}
\usepackage{fixltx2e}
\usepackage{stfloats}
\usepackage{array}
\usepackage{hyperref} 

\newcommand{\set}[1]{\mathcal{#1}}
\newcommand{\E}[1]{{\rm E}\left[{#1}\right]}
\newcommand{\Prob}[1]{{\rm Pr}\left[{#1}\right]}

\newtheorem{Theorem}{Theorem}
\newtheorem{Lemma}{Lemma}

\newtheorem{Corollary}{Corollary}
\newenvironment{proof}[1][Proof]{{\em #1:} }{}

\begin{document}

\title{Quasi-Concavity for Gaussian Multicast Relay Channels}

\newcommand{\orcidauthorA}{0000-0001-8314-500X}
\newcommand{\orcidauthorB}{0000-0002-3904-9181}

\author{Mohit Thakur and Gerhard Kramer\\
Institute for Communications Engineering, Technical University of Munich\vspace{-0.5cm}}



\maketitle
\begin{abstract}
Standard upper and lower bounds on the capacity of relay channels are cut-set (CS), decode-forward (DF), and quantize-forward (QF) rates. For real additive white Gaussian noise (AWGN) multicast relay channels with one source node and one relay node, these bounds are shown to be quasi-concave in the receiver signal-to-noise ratios and the squared source-relay correlation coefficient. Furthermore, the CS rates are shown to be quasi-concave in the relay position for a fixed correlation coefficient, and the DF rates are shown to be quasi-concave in the relay position. The latter property characterizes the optimal relay position when using DF.
\end{abstract}



\section{Introduction}

A multicast relay channel (MRC) is an information network with a source node, a relay node, and~two or more destination nodes, and where one message originating at the source should be received reliably at the destinations. We consider additive white Gaussian noise (AWGN) MRCs and show that certain information rate expressions are quasi-concave in the receiver signal-to-noise ratios (SNRs), the~squared source-relay correlation coefficient, and the relay position. In particular, we study cut-set (CS), decode-forward (DF), and quantize-forward (QF) rates. Quasi-concavity suggests that efficient algorithms can optimize signaling and the relay position. 

Relay positioning has been studied by many authors, with a focus on rate enhancement (e.g.,~\cite{Kramer-Gastpar-Gupta-2005,Lin10}), range extension (e.g.,~\cite{Aggarwal-Bennetan-Calderbank-2009,Joshi11}), and outage probability (e.g.,~\cite{Kramer-Gastpar-Gupta-2005,Lee11,Chen12}). We study the problem of placing a relay to maximize the multicast rate by extending results of~\cite{Thakur-Fawaz-Medard-Infocom2011, Thakur-Fawaz-Medard-ISIT2011, Thakur-Fawaz-Medard-ISIT2012, Thakur-Kramer-ISIT2013}. A preliminary version of this paper without proofs appeared in~\cite{Thakur-Kramer-ISIT2015}. Our focus is on real alphabet channels.
 
This paper is organized as follows. Section~\ref{sec:model} presents the MRC model and reviews the CS, DF, and QF rates. Section~\ref{sec:QRPA} develops quasi-concavity results in the squared source-relay correlation coefficient $\rho^2$ and the channel SNRs. Section~\ref{sec:qc-position} introduces a distance dependence for the channel gains and shows that the CS rate is quasi-concave in the relay position when $\rho$ is fixed. We further show that the DF rate is quasi-concave in the relay position. Section~\ref{DF-Performance} illustrates quasi-concavity for one-, two-, and three-dimensional networks, and compares the performance of two DF strategies. Section~\ref{sec:conclusions} concludes the paper. The Appendix reviews useful results on concavity and quasi-concavity, and proves a few new results.

\section{Model and Information Rates} \label{sec:model}
\vspace{-6PT}

\subsection{Model}
An MRC has three types of nodes:
\begin{itemize}
\item a source node $s$ that generates a message $W$ and transmits the symbols $X_s^n=X_{s,1},X_{s,2},\ldots,X_{s,n}$;
\item a relay node $r$ that receives and forwards symbols $Y_{r,k}$ and $X_{r,k}$, respectively, for $k=1,2, \cdots, n$;
\item destination nodes $j=1,2,\ldots,N$ where node $j$ receives $Y_j^n=Y_{j,1},Y_{j,2},\ldots,Y_{j,n}$ and estimates $W$ as~$\widehat{W}_j$.
\end{itemize}

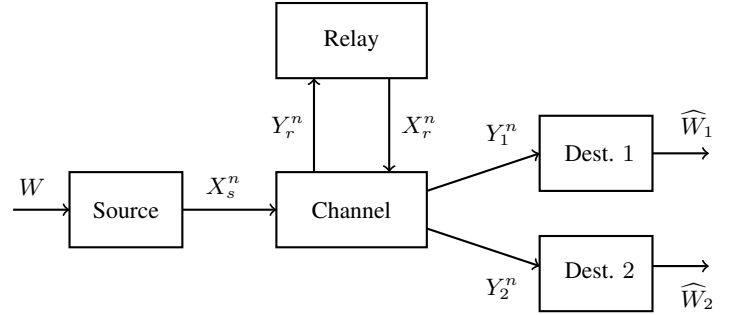
\begin{figure}[t!]
\centering
\tikzstyle{node}=[circle, draw=black, fill=black, inner sep=0pt, minimum size=3mm]
\begin{tikzpicture}
\draw [thick, ->] (-1,0.5) -- (-0.25,0.5); \node at (-0.75,0.8) {\small{$W$}};

\draw [thick] (-0.25,0) rectangle (1.25,1); \node at (0.5,0.5) {\small{Source}};
\draw [thick, ->] (1.25,0.5) -- (2.5,0.5); \node at (1.8,0.8) {\small{$X_s^{n}$}};

\draw [thick] (2.5,0) rectangle (4.5,1); \node at (3.5,0.5) {\small{Channel}};
\draw [thick] (2.5,2.25) rectangle (4.5,3.25);  \node at (3.5,2.75) {\small{Relay}};

\draw [thick, <-] (3,2.25) -- (3,1); \node at (2.65,1.6) {\small{$Y_{r}^{n}$}};
\draw [thick, ->] (4,2.25) -- (4,1); \node at (4.4,1.6) {\small{$X_{r}^{n}$}};

\draw [thick] (6,0.75) rectangle (7.5,1.75); \node at (6.8,1.25) {\small{Dest. $1$}};
\draw [thick, ->] (4.5,0.75) -- (6,1.25); \node at (5.5, 1.5) {\small{$Y^n_{1}$}};
\draw [thick, ->] (7.5,1.25) -- (8.25,1.25); \node at (8.1,1.65) {\small{$\widehat{W}_{1}$}};

\draw [thick] (6,-0.85) rectangle (7.5,0.15); \node at (6.8,-0.3) {\small{Dest. $2$}};
\draw [thick, ->] (4.5,0.25) -- (6,-0.25); \node at (5.5,-0.55) {\small{$Y^n_{2}$}};
\draw [thick, ->] (7.5,-0.25) -- (8.25,-0.25); \node at (8.1,-0.65) {\small{$\widehat{W}_{2}$}};
\end{tikzpicture}
\caption{Multicast relay channel (MRC) with two destinations.}
\label{fig:2destMRC}
\end{figure}

We denote the destination node set as $\mathcal{T}=\{1,2,\ldots,N\}$. The classic relay channel has $N=1$ and Figure~\ref{fig:2destMRC} shows an MRC with $N=2$.

A memoryless MRC has a function $h(\cdot)$ and a noise random variable ${\bf Z}$ so that for every time instant the $N+1$ channel outputs ${\bf Y}=(Y_r \; Y_1 \; \ldots \; Y_N)$ are given by
\begin{align*}
{\bf Y} = h(X_s,X_r,{\bf Z}).
\end{align*}

The noise $\bf Z$ is statistically independent of $X_s$ and $X_r$, and the noise variables at different times are statistically independent.

An encoding strategy for $M$ messages has
\begin{itemize}
\item $W$ uniformly distributed over $\{1,2,\ldots,M\}$;
\item an encoding function $e_s(\cdot)$ such that $X_s^n=e_s(W)$;
\item relay functions $e_{r,k}(\cdot)$ with $X_{r,k}=e_{r,k} (Y_{r,1},\ldots,Y_{r,k-1})$, where $k=1,2,\ldots,n$;
\item decoding functions $d_{j}(\cdot)$ such that $d_j (Y_j^n)=\widehat{W}_j $, $j\in\set{T}$.
\end{itemize}

The error probability at destination $j$ is $P_{e,j} = \Prob{\widehat{W}_j \neq W}$.
The multicast rate  is $R=(\log_2 M)/n$ bits/use.
The rate $R$ is achievable if, for any $\epsilon >0$ and sufficiently large $n$, there is an encoding strategy with $P_{e,j} \le \epsilon$ for all $j\in\set{T}$. The capacity $C$ is the supremum of the achievable rates.

\subsection{Information Rates}
The following bounds were given in \cite{Cover-ElGamal-1979} for the relay channel ($N=1$). Their extensions to MRCs are~straightforward.
\begin{itemize}
\item CS Rate: $C \le  R_{CS} $ where
\begin{align}\label{csi}
R_{CS}  = & \max \left\{ \min_{1 \le j \le N} \min \left( I(X_s X_r ; Y_j), I(X_s ; Y_r Y_{j}|X_r) \right)  \right\}
\end{align}
and where the maximization is over all $X_s X_r$.
\item Direct-Transmission (DT) Rate: $C \ge  R_{DT}$ where
\begin{align}\label{dti}
R_{DT} = \max \{ \displaystyle\min_{1 \le j \le N} I(X_s; Y_{j}|X_{r}= x^*) \}
\end{align}
and where the maximization is over all $x^*$ and $X_s$.
\item DF Rate: $C \ge  R_{DF}$ where
\begin{align}\label{dfi}
R_{DF} = & \max \left\{ \min_{1 \le j \le N} \min \left( I(X_s X_r;Y_j), I(X_s;Y_r | X_r) \right)  \right\}
\end{align}
and where the maximization is over all $X_s X_r$.
\item QF Rate: $C \ge  R_{QF}$ where
\begin{align}\label{qfi}
R_{QF} = & \max \left\{ \min_{1 \le j \le N} \min \left( I(X_s X_r; Y_j) - \right. \right. \nonumber \\
& \quad \left. \left. I(Y_r; \widehat{Y}_r | X_s X_r Y_j), I(X_s;\widehat{Y}_rY_j|X_r) \right) \right\}
\end{align}
where $\widehat{Y}_r$ is an auxiliary random variable,
and where the maximization is over all $X_s X_r \widehat{Y}_r$ such that
$X_s$ and $X_r$ are independent and $X_s - X_r Y_r - \widehat{Y}_r$ forms a Markov chain.
\end{itemize}

\subsection{Real Alphabet AWGN MRC}

\begin{figure}[t!]
\centering
\tikzstyle{node}=[circle, draw=black, fill=black, inner sep=0pt, minimum size=2mm]
\begin{tikzpicture}
\node at (0,0) (s) [node, label=below:$s$] {};
\node at (2,0) (r) [node, label=below:$r$] {};
\node at (4,2) (t1) [node, label=below:$1$] {};
\node at (4,-2) (t2) [node, label=above:$2$] {};

\draw [thick, dashed] (s) -- (r);
\draw [thick, dashed] (s) -- (t1); \draw [thick, dashed] (s) -- (t2);
\draw [thick, dashed] (r) -- (t1); \draw [thick, dashed] (r) -- (t2);

\node at (1.1,0.2) {$a_{s,r}$}; \node at (2,1.3) {$a_{s,1}$}; \node at (2,-1.3) {$a_{s,2}$};
\node at (3.2,0.6) {$a_{r,1}$}; \node at (3.2,-0.7) {$a_{r,2}$};
\end{tikzpicture}
\caption{AWGN MRC with two destinations.}
\label{fig:GMRC2}
\end{figure}
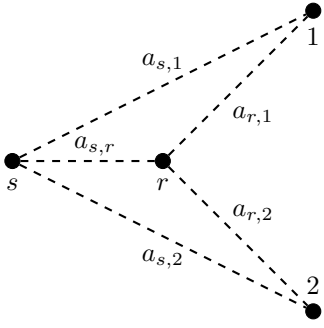

The real alphabet AWGN MRC has real channel symbols and
\begin{align}
Y_{r} & = a_{s,r} X_s + Z_{r} \label{eq:Yr} \\
Y_{j} & = a_{s,j} X_s + a_{r,j} X_{r} + Z_{j} \label{eq:Yj}
\end{align}
where $j \in \mathcal{T}$. The $a_{s,r}$, $a_{s,j}$, and $a_{r,j}$ are channel
gains between the nodes (see Figure~\ref{fig:GMRC2}). We later relate these
gains to distances between the nodes.
The $Z_{r}$ and $Z_{j}$, $j=1,2,\ldots,N$, are independent and identically
distributed Gaussian random variables with zero mean and unit variance.
We may alternatively write \eqref{eq:Yr} and \eqref{eq:Yj} in vector form as
\begin{align}
\mathbf{Y}_j = \mathbf{A}_j \mathbf{X} + \mathbf{Z}_j
\end{align}
where $\mathbf{X}=(X_s \; X_r)^T$, $\mathbf{Y}_j=(Y_r \; Y_j)^T$, $\mathbf{Z}=(Z_r \; Z_j)^T$, and
\begin{align}
\mathbf{A}_j = \begin{pmatrix} a_{s,r} & 0 \\ a_{s,j} & a_{r,j} \end{pmatrix}.
\end{align}

We consider individual average block power constraints
\begin{align}\label{powerconstraints}
\E{\sum_{k=1}^{n} X^2_{s,k}} \le n P_s, \quad
\E{\sum_{k=1}^{n} X^2_{r,k}} \le n P_r.
\end{align}

The SNR and the capacity of the link from node $u$ (with transmit power $P_u$) to node $v$ are the respective
\begin{align}
& \mathsf{SNR}_{u,v}= a_{u,v}^2 P_u  \label{eq:SNR} \\
& \mathsf{C}(\mathsf{SNR}_{u,v})=\frac{1}{2} \log \left( 1 + \mathsf{SNR}_{u,v} \right).
\end{align}

We simplify the above rate bounds for the AWGN MRC.
\begin{itemize}
\item CS Rate:
\begin{align}\label{cs}
R_{CS} & = \max_{\rho} \left[ \displaystyle\min_{1 \le j \le N} \min  \left( \right. \right. \nonumber \\
& \qquad \mathsf{C}\left(\mathsf{SNR}_{s,j} + \mathsf{SNR}_{r,j} + 2 \rho \sqrt{\mathsf{SNR}_{s,j} \mathsf{SNR}_{r,j}} \right),  \nonumber \\
& \qquad \left. \mathsf{C}\left((1 - \rho^2) (\mathsf{SNR}_{s,j} + \mathsf{SNR}_{s,r})) \right)\right]
\end{align}
where the correlation coefficient $\rho$ satisfies $|\rho|\le1$. One can restrict attention to non-negative $\rho$.
\item DT Rate:
\begin{align}\label{dt}
R_{DT} = \min_{1 \le j \le N} \mathsf{C}(\mathsf{SNR}_{s,j}).
\end{align}
\item DF Rate:
\begin{align}\label{df}
& R_{DF} = \max_{\rho} \left[ \min_{1 \le j \le N} \min \left( \right. \right. \nonumber \\
& \qquad \mathsf{C}(\mathsf{SNR}_{s,j}  + \mathsf{SNR}_{r,j} + 2 \rho \sqrt{\mathsf{SNR}_{s,j} \mathsf{SNR}_{r,j}} ), \nonumber \\
& \qquad  \left.  \left.  \mathsf{C}((1 - \rho^2) \mathsf{SNR}_{s,r} ) \right) \right].
\end{align}
One can again restrict attention to non-negative $\rho$.
\item QF Rate:
Optimizing $X_sX_r\widehat{Y}_r$ seems difficult. Instead, we choose $X_s$ and $X_r$ to be zero-mean
Gaussian with variances $P_s$ and $P_r$, respectively. We further choose
$\widehat{Y}_r = Y_{r}+Z_r$ where $Z_r$ is zero-mean Gaussian with variance $N_r$.
Optimizing ${N}_{r}$ gives (see~\cite{Kramer-Maric-Yates-2006}, pp.~336--337)
\begin{align}\label{qf}
\widetilde{R}_{QF} = & \min_{1 \le j \le N} \mathsf{C} \left( \mathsf{SNR}_{s,j} +
\frac{\mathsf{SNR}_{r,j} \mathsf{SNR}_{s,r}}{\mathsf{SNR}_{s,j} + \mathsf{SNR}_{r,j} + \mathsf{SNR}_{s,r} +1} \right).
\end{align}
\end{itemize}

\section{Quasi-Concavity in SNRs and $\rho^2$}\label{sec:QRPA}
\vspace{-6PT}

\subsection{CS Rate}

We consider two characterizations of $R_{CS}$. First, let $\mathbf{a}_j^T=(a_{s,j} \; a_{r,j})$
be the second row of $\mathbf{A}_j$, let $\mathbf{Q}_{\mathbf{X}}$ be the covariance
matrix of $\mathbf{X}$ (see Appendix~\ref{sec:cov-con}), and let $\det \mathbf{M}$ be
the determinant of the square matrix $\mathbf{M}$. The CS rate \eqref{cs} can
be expressed as the maximum of
\begin{align}\label{CS-1}
R_{CS}(\mathbf{Q}_{\mathbf{X}}) = & \min_{1 \le j \le N} \min  \left( \frac{1}{2}\log\left( \mathbf{a}_j^T \: \mathbf{Q}_{\mathbf{X}} \: \mathbf{a}_j +1 \right), \right. \nonumber \\
& \qquad \left. \frac{1}{2}\log\left( \frac{\det \mathbf{Q}_{(\mathbf{Y}_j^T \; X_r )^T}}{P_r} \right) \right)
\end{align}
over the convex set of $\mathbf{Q}_{\mathbf{X}}$ with diagonal entries $P_s$ and $P_r$. The first logarithm in \eqref{CS-1} is clearly concave in $\mathbf{Q}_{\mathbf{X}}$.
The second logarithm is concave in $\mathbf{Q}_{(\mathbf{Y}_j^T \; X_r )^T}$ (see Appendix~\ref{sec:cov-con}) and
$\mathbf{Q}_{(\mathbf{Y}_j^T \; X_r )^T}$ is linear in $\mathbf{Q}_{\mathbf{X}}$. To~prove the latter claim, observe that
\begin{align}
\mathbf{Q}_{(\mathbf{Y}_j^T \; X_r )^T}
= \tilde{\mathbf{A}}_j \mathbf{Q}_{\mathbf{X}} \tilde{\mathbf{A}}_j^T
+ \left( \begin{array}{cc} \mathbf{I}_{2} & \mathbf{0} \\ \mathbf{0} & 0 \end{array} \right)
\end{align}
where $\tilde{\mathbf{A}}_j^T=\left( \mathbf{A}_j^T \; \left[ 0 \; 1 \right]^T \right)$ and $\mathbf{I}_{2}$ is the $2 \times 2$ identity matrix. Hence $R_{CS}(\mathbf{Q}_{\mathbf{X}})$ is concave in (the convex set of) $\mathbf{Q}_{\mathbf{X}}$ because it is the minimum of $2N$ concave functions.

Suppose next that we wish to consider $\rho$ and the SNRs
individually rather than via $\mathbf{Q}_{\mathbf{X}}$. Define the~vector
\begin{align}
\mathbf{S} & =(\mathsf{SNR}_{s,r},\mathsf{SNR}_{s,1}, \cdots, \mathsf{SNR}_{s,N}, \mathsf{SNR}_{r,1}, \cdots, \mathsf{SNR}_{r,N})
\end{align}
and the functions
\begin{align}
& f_j(\rho,\mathbf{S}) = \mathsf{SNR}_{s,j} + \mathsf{SNR}_{r,j} + 2 \rho \sqrt{\mathsf{SNR}_{s,j} \mathsf{SNR}_{r,j}} \label{eq:fj} \\
& g_j(\rho,\mathbf{S}) = (1-\rho^2)\left( \mathsf{SNR}_{s,j} + \mathsf{SNR}_{s,r}\right) \label{eq:gj} \\
& R_{CS}(\rho,\mathbf{S}) = \min_{1 \le j \le N} \min  \left(\mathsf{C}(f_j(\rho,\mathbf{S})),\mathsf{C}(g_j(\rho,\mathbf{S})) \right) .
\label{eq:RCS}
\end{align}

We establish the following results. We restrict attention to $0 \le \rho \le 1$ and positive $\mathbf{S}$.

\begin{Lemma}\label{lemma:CS}
$f_j(\rho,\mathbf{S})$ and $g_j(\rho,\mathbf{S})$ are concave in $\rho$,
concave in $\mathbf{S}$, and quasi-concave in $(\rho^2,\mathbf{S})$.
\end{Lemma}
\begin{proof}
Concavity with respect to $\rho$ is established by observing that $f_j(\rho,\mathbf{S})$ is
linear in $\rho$, and $g_j(\rho,\mathbf{S})$ is linear in $-\rho^2$ which is concave in $\rho$.

Consider next concavity with respect to $\mathbf{S}$.
The Hessian of $f_j(\rho,\mathbf{S})$ with respect to $\mathbf{S}$ has only one non-zero eigenvalue
\begin{align}
- \frac{\rho}{2} \cdot \frac{\mathsf{SNR}_{s,j}^2 + \mathsf{SNR}_{r,j}^2}{\mathsf{SNR}_{s,j}^{3/2} \mathsf{SNR}_{r,j}^{3/2}}.
\end{align}
Thus, $f_j(\rho,\mathbf{S})$ is concave in $\mathbf{S}$ for non-negative $\rho$ and positive $\mathbf{S}$.
The function $g_j(\rho,\mathbf{S})$ is linear in $\mathbf{S}$, and~thus concave in $\mathbf{S}$.

Now consider quasi-concavity with respect to $(\rho^2,\mathbf{S})$.
Substituting $a=\mathsf{SNR}_{s,j},b=\mathsf{SNR}_{r,j},c=\rho^2$ into the fifth function of Lemma~\ref{lemma:qcfunc1}
{in Appendix}~\ref{sec:CA}, we find that $f_j(\rho,\mathbf{S})$ is quasi-concave in $(\rho^2,\mathbf{S})$.
For the $g_j(\rho,\mathbf{S})$, observe that $ab$ is quasi-concave for non-negative $(a,b)$,
see the first function of Lemma~\ref{lemma:qcfunc1}. This~implies
\begin{align}
(\lambda a_1+\bar{\lambda} a_2)(\lambda b_1+\bar{\lambda} b_2)
\ge \min\left( a_1 b_1, a_2 b_2 \right)
\end{align}
for $0\le\lambda\le1$, and where $\bar{\lambda}=1-\lambda$. Substituting
$a_i=1-\rho_i^2$ and $b_i=\mathsf{SNR}_{s,j,i}+\mathsf{SNR}_{s,r,i}$ for $i=1,2$, we~find that
$g_j(\rho,\mathbf{S})$ is quasi-concave in $(\rho^2,\mathbf{S})$.
\end{proof}
\begin{Theorem}\label{thm:CS}
$R_{CS}(\rho,\mathbf{S})$ is concave in $\rho$,
concave in $\mathbf{S}$,
and quasi-concave in $(\rho^2,\mathbf{S})$.
\end{Theorem}
\begin{proof}
$R_{CS}(\rho,\mathbf{S})$ involves taking logarithms and minima of (quasi-) concave functions.
The results thus follow by applying Lemma~\ref{lemma:CS} {above} and
Lemma~\ref{lemma:qcprops}, Parts~\ref{Part-min} and~\ref{Part-comp}, {in Appendix}~\ref{sec:CA}.
\end{proof}
\begin{Corollary}\label{cor:CS}
Consider $\mathbf{S}$ as a function of $\mathbf{P}=(P_s,P_r)$.
Then $R_{CS}(\rho,\mathbf{S}(\mathbf{P}))$ is quasi-concave in $(\rho^2,\mathbf{P})$.
\end{Corollary}
\begin{proof}
The proof follows from the proof of Theorem~\ref{thm:CS} and because
$\mathbf{S}$ is a linear function of $\mathbf{P}$.
\end{proof}
%

\subsection{DF Rate}
Consider the functions
\begin{align}
& g^*_j(\rho,\mathbf{S}) = (1-\rho^2) \mathsf{SNR}_{s,r} \label{eq:gtj} \\
& R_{DF}(\rho,\mathbf{S}) = \min_{1 \le j \le N} \min  \left(\mathsf{C}(f_j(\rho,\mathbf{S})),  \mathsf{C}(g^*_j(\rho,\mathbf{S})) \right).  \label{eq:DF}
\end{align}
As above, we restrict attention to $0 \le \rho \le 1$ and positive $\mathbf{S}$.

\begin{Theorem}\label{thm:DF}
$R_{DF}(\rho,\mathbf{S})$ is concave in $\rho$,
concave in $\mathbf{S}$, and quasi-concave in $(\rho^2,\mathbf{S})$.
\end{Theorem}
\begin{proof}
The proof is similar to that of Theorem~\ref{thm:CS}.
\end{proof}
\begin{Corollary}\label{cor:DF}
$R_{DF}(\rho,\mathbf{S}(\mathbf{P}))$ is quasi-concave in $(\rho^2,\mathbf{P})$.
\end{Corollary}
\begin{proof}
See the proof of Corollary~\ref{cor:CS}.
\end{proof}
%

\subsection{DT Rate}
%
The DT rate \eqref{dt} is clearly concave in $\mathbf{S}$ and $\mathbf{P}$.

\subsection{QF Rate}

Consider the functions
\begin{align}
& h_j(\mathbf{S}) = \mathsf{SNR}_{s,j} + \frac{\mathsf{SNR}_{r,j} \mathsf{SNR}_{s,r}}{\mathsf{SNR}_{s,j} + \mathsf{SNR}_{r,j} + \mathsf{SNR}_{s,r} +1}
\label{eq:hj} \\
& \widetilde{R}_{QF}(\mathbf{S}) = \min_{1 \le j \le N} \mathsf{C}(h_j(\mathbf{S})). \label{eq:RQF}
\end{align}
We establish the following results. We restrict attention to non-negative $\mathbf{S}$.
\begin{Lemma}\label{lemma:QF}
$h_j(\mathbf{S})$ is quasi-concave in $(\mathsf{SNR}_{r,j},\mathsf{SNR}_{s,r})$.
\end{Lemma}
\begin{proof}
Substitute $a=\mathsf{SNR}_{r,j},b=\mathsf{SNR}_{s,r},k=\mathsf{SNR}_{s,j} +1$ into the second function of
Lemma~\ref{lemma:qcfunc1} {in Appendix}~\ref{sec:CA},
and apply Lemma~\ref{lemma:qcprops}, Part~\ref{Part-lin}.
\end{proof}
\begin{Theorem}\label{QF}
$\widetilde{R}_{QF}(\mathbf{S})$ is quasi-concave in $\mathbf{S}$ if the $\mathsf{SNR}_{s,j}$,
$j=1,2,\ldots,n$, are held fixed.
\end{Theorem}
\begin{proof}
Apply {Lemma}~\ref{lemma:QF} {above} and Lemma~\ref{lemma:qcprops}, Parts~\ref{Part-min}
and~\ref{Part-comp}, {in Appendix}~\ref{sec:CA}.
\end{proof}

\section{Quasi-Concavity in Relay Position} \label{sec:qc-position}

Suppose the channel gain for the node pair $(i,j)$ is
\begin{align}
a_{i,j} = \sqrt{\xi_{i,j}} \left/ D_{i,j}^{\alpha/2} \right. \label{eq:aij}
\end{align}
where $\xi_{i,j}$ is a ``fading'' gain, $D_{i,j} = \|\mathbf{i} -\mathbf{j}\|$ is the Euclidean
distance between the positions $\mathbf{i}$ and $\mathbf{j}$ of nodes $i$ and $j$,
respectively, and $\alpha\ge2$ is a path-loss exponent. We thus have
\begin{align*}
\mathsf{SNR}_{i,j} =\frac{\xi_{i,j} P_i}{D_{i,j}^{\alpha}} = \frac{\xi_{i,j} P_i}{\|\mathbf{i} -\mathbf{j} \|^{\alpha}}.
\end{align*}
We establish quasi-concavity results in $\rho^2$ and $\mathbf{r}$, where $\mathbf{r}$ is the position of the relay node.

\subsection{CS Rate}

Consider the functions \eqref{eq:fj}--\eqref{eq:RCS} but relabeled as
$f_j(\rho,\mathbf{r})$, $g_j(\rho,\mathbf{r})$, and $R_{CS}(\rho,\mathbf{r})$
to emphasize the dependence on the considered parameters.
We again consider $0 \le \rho \le 1$ and positive $\mathbf{S}$.
\begin{Lemma}\label{lemma:LOSCS}
$f_j(\rho,\mathbf{r})$ and $g_j(\rho,\mathbf{r})$ are quasi-concave in $\mathbf{r}$ for fixed $\rho$.
Furthermore, $f_j(\rho,\mathbf{r})$ is quasi-concave in $(\rho^2,\mathbf{r})$.
\end{Lemma}
\begin{proof}
Consider the functions
\begin{align}
& \tilde{f}_j(\rho,D^{\alpha}) =
\frac{\xi_{s,j} P_{s}}{D^{\alpha}_{s,j}} + \frac{\xi_{r,j} P_{r}}{D^{\alpha}}
+ 2 \rho \sqrt{\frac{\xi_{s,j} P_{s}}{D^{\alpha}_{s,j}}\frac{\xi_{r,j} P_{r}}{D^{\alpha}}} \label{eq:CS-fjt} \\
& \tilde{g}_j(\rho,D^{\alpha}) = (1- \rho^2) \left(\frac{\xi_{s,j} P_{s}}{D_{s,j}^{\alpha}}
+ \frac{\xi_{s,r} P_{s}}{D^{\alpha}} \right) \label{eq:CS-gjt}
\end{align}
which are quasi-linear in $D^{\alpha}$ for fixed $\rho$ since they are decreasing in $D^\alpha$.
However, $D_{r,j}^\alpha$ is a convex function of $\mathbf{r}$ for $\alpha\ge1$, and thus Lemma~\ref{lemma:qcprops},
Part~\ref{Part-con}, {in Appendix}~\ref{sec:CA} establishes that $f_j(\rho,\mathbf{r})$ is quasi-concave in $\mathbf{r}$ for fixed $\rho$.
Similarly, $D_{s,r}^\alpha$ is a convex function of $\mathbf{r}$ for $\alpha\ge1$, and we find that
$g_j(\rho,\mathbf{r})$ is quasi-concave in $\mathbf{r}$ for fixed $\rho$.

Next, substitute $a=D^\alpha$ and $b=\rho^2$ into the third function of Lemma~\ref{lemma:qcfunc1},
and use Lemma~\ref{lemma:qcprops}, Part~\ref{Part-lin}, to show that $\tilde{f}_j(\rho,D^\alpha)$ is
quasi-concave in $(\rho^2,D^\alpha)$. However, $\tilde{f}_j$ is decreasing in $D^\alpha$ and
$D_{r,j}^\alpha$ is convex in $\mathbf{r}$, so Lemma~\ref{lemma:qcprops}, Part~\ref{Part-con},
establishes that $f_j(\rho,\mathbf{r})$ is quasi-concave in $(\rho^2,\mathbf{r})$.
\end{proof}

Unfortunately, $\tilde{g}_j$ is quasi-{\em convex} (and not
quasi-concave) in $(\rho^2,D^{\alpha})$. To see this, substitute
$a=D^\alpha$ and $b=\rho^2$ into the fourth function of Lemma~\ref{lemma:qcfunc1}.
Quasi-concavity would have been useful since it would have permitted using
Lemma~\ref{lemma:qcprops}, Parts~\ref{Part-min} and~\ref{Part-sup}, to establish the quasi-concavity of
\begin{align}
R_{CS}(\mathbf{r}) = \max_{\rho} \left[ \min_{1 \le j \le N} \min
\left(\mathsf{C}(f_j(\rho,\mathbf{r})),\mathsf{C}(g_j(\rho,\mathbf{r})) \right) \right].
\label{eq:RCSr}
\end{align}
However, we have been unable to prove this, and our numerical results suggest that
$R_{CS}(\rho,\mathbf{r})$ is not quasi-concave in $(\rho^2,\mathbf{r})$.
Nevertheless, Lemma~\ref{lemma:LOSCS} suffices to
establish an intermediate result which is useful in Section~\ref{DF-Performance} when we study $\rho=0$.
\begin{Theorem}\label{thm:LOSCS}
$R_{CS}(\rho,\mathbf{r})$ is quasi-concave in $\mathbf{r}$ for fixed $\rho$, $0\le\rho\le1$.
\end{Theorem}
\begin{proof}
$R_{CS}(\rho,\mathbf{r})$ is the minimum of functions
that are quasi-concave in $\mathbf{r}$. Lemma~\ref{lemma:qcprops}, Part~\ref{Part-min},
thus~establishes the theorem.
\end{proof}

\subsection{DF Rate}
%
The quasi-convexity of $\tilde{g}_j(\rho,D^\alpha)$ relaxes for the DF rate~\eqref{eq:DF}.
Consider the negative of the fourth function of Lemma~\ref{lemma:qcfunc1} {in Appendix}~\ref{sec:CA}
with $k_1=0$:
\begin{align}
f(a,b) = (1-b) k_2 / a .
\end{align}
{This} function is quasi-linear in $(a,b)$ since both its superlevel and sublevel sets
are convex. This result implies the following theorem.
We again consider the functions \eqref{eq:gtj}--\eqref{eq:DF} but relabeled as
$g^*_j(\rho,\mathbf{r})$ and $R_{DF}(\rho,\mathbf{r})$. We further define
\begin{align}
& \tilde{g}^*_j(\rho,D^{\alpha}) = (1- \rho^2) \frac{\xi_{s,r} P_{s}}{D^{\alpha}} \\
& R_{DF}(\mathbf{r}) = \max_{\rho} \left[ \min_{1 \le j \le N} \min
\left(\mathsf{C}(f_j(\rho,\mathbf{r})),\mathsf{C}(g^*_j(\rho,\mathbf{r})) \right) \right].
\label{eq:RDFr}
\end{align}
As above, we consider $0 \le \rho \le 1$ and positive $\mathbf{S}$.

\begin{Theorem}\label{thm:LOSDF}
$R_{DF}(\rho,\mathbf{r})$ is quasi-concave in $(\rho^2,\mathbf{r})$,
and $R_{DF}(\mathbf{r})$ is quasi-concave in $\mathbf{r}$.
\end{Theorem}
\begin{proof}
$\tilde{g}^*_j(\rho,D^{\alpha})$ is quasi-linear in $(\rho^2,D^{\alpha})$
and decreasing in $D^\alpha$. Furthermore, $D_{s,r}^\alpha$ is convex in
$\mathbf{r}$, and thus Lemma~\ref{lemma:qcprops}, Part~\ref{Part-con}, {in Appendix}~\ref{sec:CA}
establishes that $g^*_j(\rho,\mathbf{r})$ is quasi-concave in $(\rho^2,\mathbf{r})$.
$R_{DF}(\rho,\mathbf{r})$ is therefore quasi-concave in $\mathbf{r}$, as it is the minimum of
quasi-concave functions (see Lemma~\ref{lemma:qcprops}, Part~\ref{Part-min}).
Furthermore, $R_{DF}(\mathbf{r})$ is concave in $\mathbf{r}$ by
Lemma~\ref{lemma:qcprops}, Part~\ref{Part-sup}.
\end{proof}

\section{DF Performance}\label{DF-Performance}

This section presents numerical results for the DF strategy and compares them to results from~\cite{Thakur-Fawaz-Medard-Infocom2011,Thakur-Fawaz-Medard-ISIT2011,Thakur-Fawaz-Medard-ISIT2012}.
We consider 1-, 2-, and 3-dimensional MRCs with different numbers $N$ of destination nodes.
For simplicity, we consider the low SNR or broadband regime where
\begin{align}
\mathsf{C}(\mathsf{SNR}) = \frac{1}{2} \log(1+\mathsf{SNR}) \rightarrow \frac{1}{2} \mathsf{SNR}.
\end{align}

In other words, we consider the CS and DF rates without the logarithms. This approach
is valid not only in the limit of low SNR, but more generally because we proved our quasi-concavity
results without taking logarithms. Furthermore, in the low SNR regime the rates of
full-duplex and half-duplex transmission are the same under a block power constraint.

We choose $P_{s}=P_{r}=P=1$, $\alpha=2$, and $\xi_{u,v}=1$ for all node pairs $(u,v)$.
We study both coherent transmission where $\rho$ is optimized and non-coherent
transmission with $\rho=0$. The rates are in nats/channel use. Alternatively, suppose
we use sync pulses sampled at $2W$ samples per second, where $W$ is the (one-sided)
signal bandwidth. Suppose further that the (one-sided) noise power spectral density is
1~Watt/Hz. Then at low SNR the rates in nats/channel use are the same as the rates in nats/sec.

\subsection{One Dimension}\label{subsec:oned}

\begin{figure}[t!]
\centering
\includegraphics[width=0.95\columnwidth]{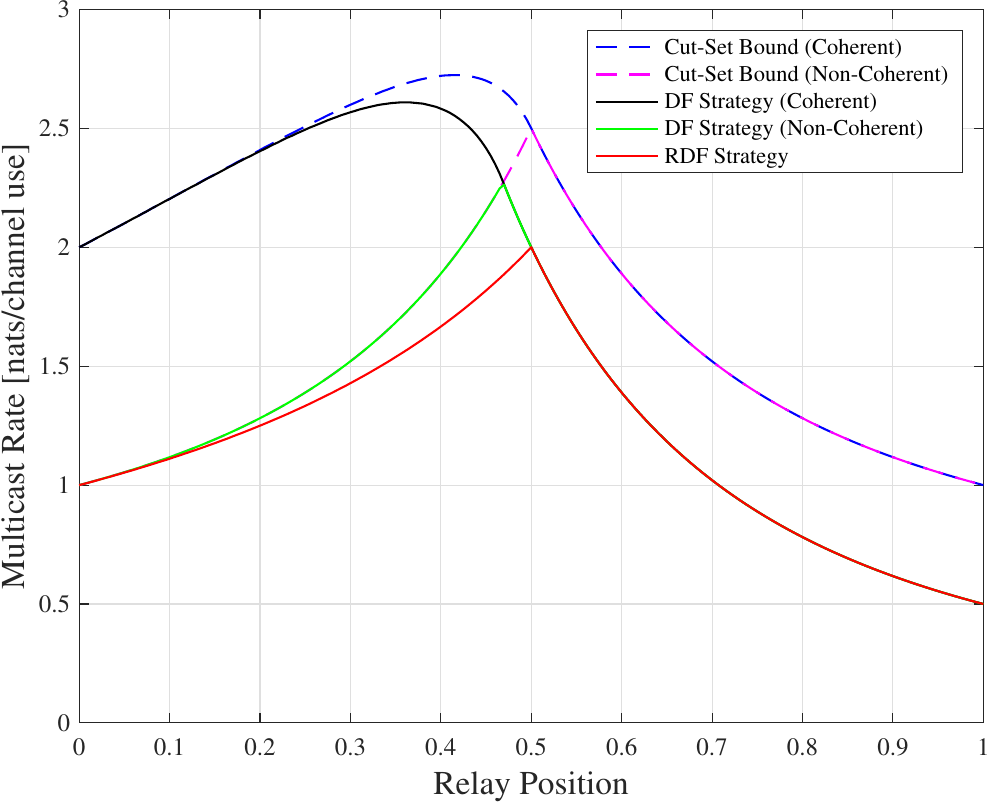}
\caption{Relay channel rates for low signal-to-noise ratio (SNR) and $P=1$.}
\label{fig:AllPerfComp}
\end{figure}

Consider a relay channel ($N=1$) where the source is at the origin ($\mathbf{s}=0$)
and the destination is at point 1 ($\mathbf{1}=1$).
Figure~\ref{fig:AllPerfComp} shows the low SNR CS rates, DF rates, and
the routing-based DF (RDF) rates developed in~\cite{Thakur-Fawaz-Medard-Infocom2011}.
Observe that all curves are quasi-concave (but not concave) in $\mathbf{r}$.
Theorems~\ref{thm:LOSCS} and~\ref{thm:LOSDF} predict the quasi-concavity for all curves except for the coherent CS rates.
Observe also that the curves for the coherent and non-coherent rates merge for relay positions exceeding a certain value
($r=0.5$ and $r\approx0.47$ for the respective CS and DF rates).
The reason for this behavior is that $\rho=0$ is optimal for the coherent CS and DF rates beyond these positions,
see the $\rho$ curve in~\cite{Kramer-Gastpar-Gupta-2005} (Figure~16). Furthermore, the non-coherent CS rates
coincide with {the non-coherent DF rates} for a large range of $\mathbf{r}$.

The best relay positions for the two strategies are different. For example, $\mathbf{r}=0.5$ maximizes $R_{RDF}$ while the $\mathbf{r}$ maximizing $R_{DF}$ is closer to the source. This is because when the source transmits, the relay and the destination {listen}, and the destination ``collects'' information. The relay can thus be positioned closer to the source while maintaining the same information rate from the source to the relay, and from the source-relay pair to the destination. At the optimal positions, we compute $R_{DF} \approx 2.26 P$ nats/sec and $R_{RDF}=2 P$ nats/sec, so the DF gain is $\approx$13\%.

\subsection{Two Dimensions}\label{subsec:twod}
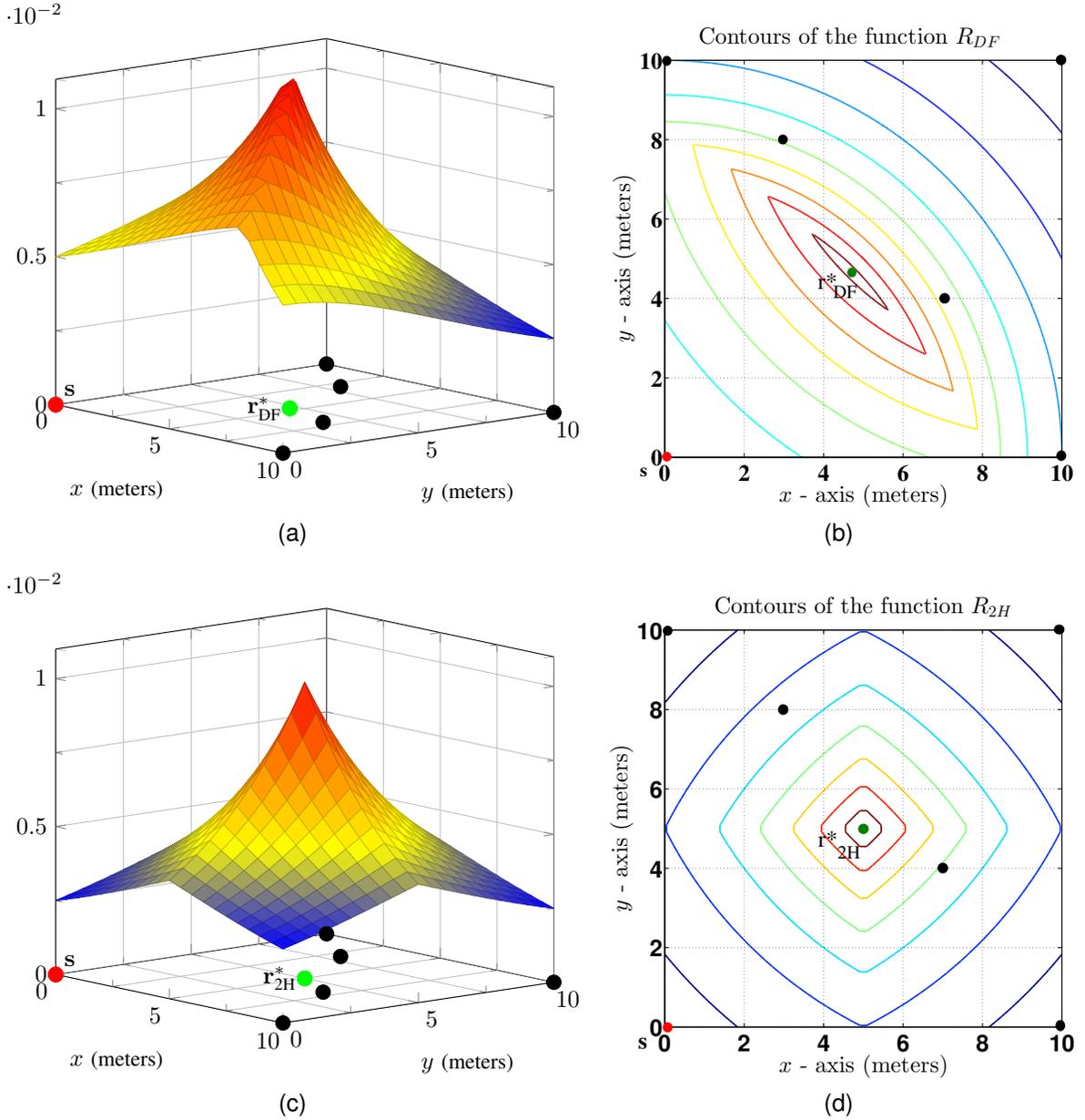
\begin{figure*}[t!]
\centering
\newcommand*\thesubfloatfigure{\themainfigure\Alph{subfloatfigure}}
\subfloat[Subfigure 1][]{
\pgfplotsset{width=1.0\columnwidth,compat=1.8}
\begin{tikzpicture}
[
declare function={
r1(\x,\y) =0.5 * min(1/200 + 1/((\x-10)^2 + (\y-10)^2),1/((\x)^2 + (\y)^2));
r2(\x,\y) =0.5 * min(1/100 + 1/((\x-10)^2 + (\y)^2),1/((\x)^2 + (\y)^2));
r3(\x,\y) =0.5 * min(1/100 + 1/((\x)^2 + (\y -10)^2),1/((\x)^2 + (\y)^2));
r4(\x,\y) =0.5 * min(1/65 + 1/((\x - 7)^2 + (\y -4)^2),1/((\x)^2 + (\y)^2));
r5(\x,\y) =0.5 * min(1/73 + 1/((\x - 3)^2 + (\y -8)^2),1/((\x)^2 + (\y)^2));
}, scale=1
]
\begin{axis}[view={50}{11}, grid=both, minor tick num=1, xtick={0, 5, 10}, ytick={0, 5, 10},
xlabel=$\small x \text{ (meters)}$, ylabel=$\small y \text{ (meters)}$,
enlarge z limits=0, zmin=0, zmax=0.011,
]

\addplot3[surf, samples=21, domain = 0:10, domain y=0:10] 
{min(r1(x,y),r2(x,y),r3(x,y),r4(x,y),r5(x,y))};

\addplot+[mark=*, red, mark options={scale=1.5, fill=red}] table[only marks, x=sx,y=sy] {nodes.dat};
\addplot [mark=*, black, only marks, mark options={scale=1.5, fill=black}] table[only marks, x=x,y=y] {nodes.dat};
\addplot [mark=*, green, mark options={scale=1.5, fill=green}] table[only marks, x=r1x,y=r1y] {nodes.dat};

\node at (0,0) [above right,font=\normalsize] {$\mathbf{s}$};
\node at (axis cs:4.7,4.7,0) [left, font=\normalsize] {$\mathbf{r}^*_{\text{DF}}$};

\end{axis}
\end{tikzpicture}
\label{fig:FivedestNCDF}
} 
%
\subfloat[Subfigure 2][]{\includegraphics[width=0.8\columnwidth]{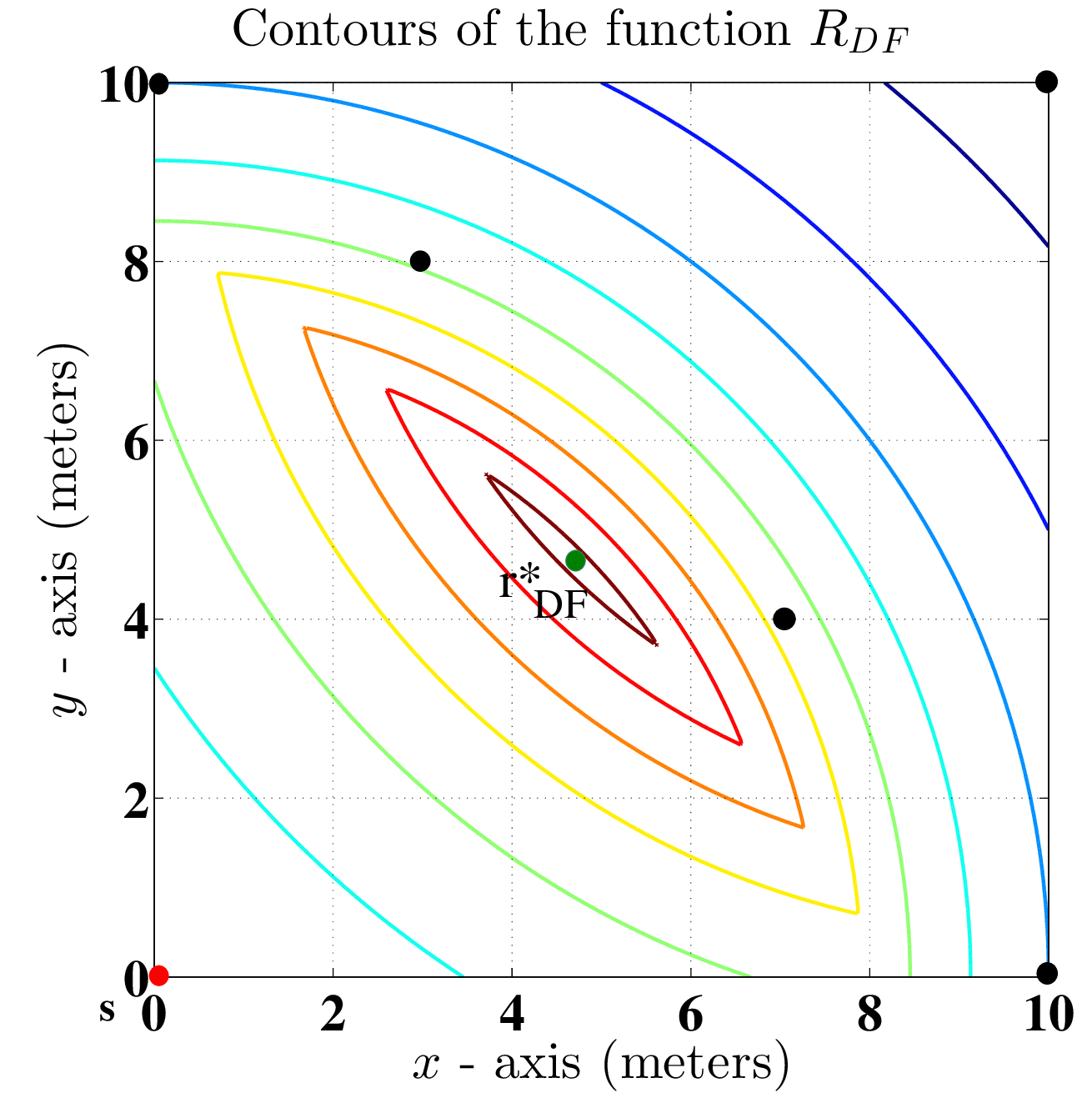}\label{fig:NCDFContour}}
\\
%
\subfloat[Subfigure 3][]{
\pgfplotsset{width=1.0\columnwidth,compat=1.8}
\begin{tikzpicture}
[
declare function={
r1(\x,\y) = 0.5 * min(1/((\x-10)^2 + (\y-10)^2),1/((\x)^2 + (\y)^2));
r2(\x,\y) = 0.5 * min(1/((\x-10)^2 + (\y)^2),1/((\x)^2 + (\y)^2));
r3(\x,\y) = 0.5 * min(1/((\x)^2 + (\y -10)^2),1/((\x)^2 + (\y)^2));
r4(\x,\y) = 0.5 * min(1/((\x - 7)^2 + (\y -4)^2),1/((\x)^2 + (\y)^2));
r5(\x,\y) = 0.5 * min(1/((\x - 3)^2 + (\y -8)^2),1/((\x)^2 + (\y)^2));
}, scale=1
]
\begin{axis}[view={50}{11}, grid=both, minor tick num=1, xtick={0, 5, 10}, ytick={0, 5, 10},
xlabel=$\small x \text{ (meters)}$, ylabel=$\small y \text{ (meters)}$,
enlarge z limits=0, zmin=0, zmax=0.011,
]
ax=gca;
ax.XTick=[0,5,10];
ax.YTick=[0,5,10];

\addplot3[surf, samples=21, domain = 0:10, domain y=0:10,] 
{min(r1(x,y),r2(x,y),r3(x,y),r4(x,y),r5(x,y))};
\addplot+[mark=*, red, mark options={scale=1.5, fill=red}] table[only marks, x=sx,y=sy] {nodes.dat};
\addplot [mark=*, black, only marks, mark options={scale=1.5, fill=black}] table[only marks, x=x,y=y] {nodes.dat};
\addplot [mark=*, green, mark options={scale=1.5, fill=green}] table[only marks, x=r2x,y=r2y] {nodes.dat};

\node at (axis cs:0,0,0) [above right,font=\normalsize] {$\mathbf{s}$};
\node at (axis cs:5,5,0) [left, font=\normalsize] {$\mathbf{r}^*_{\text{2H}}$};

\end{axis}
\label{fig:FivedestRDF}
\end{tikzpicture}
}
%
\subfloat[Subfigure 4][]{\includegraphics[width=0.8\columnwidth]{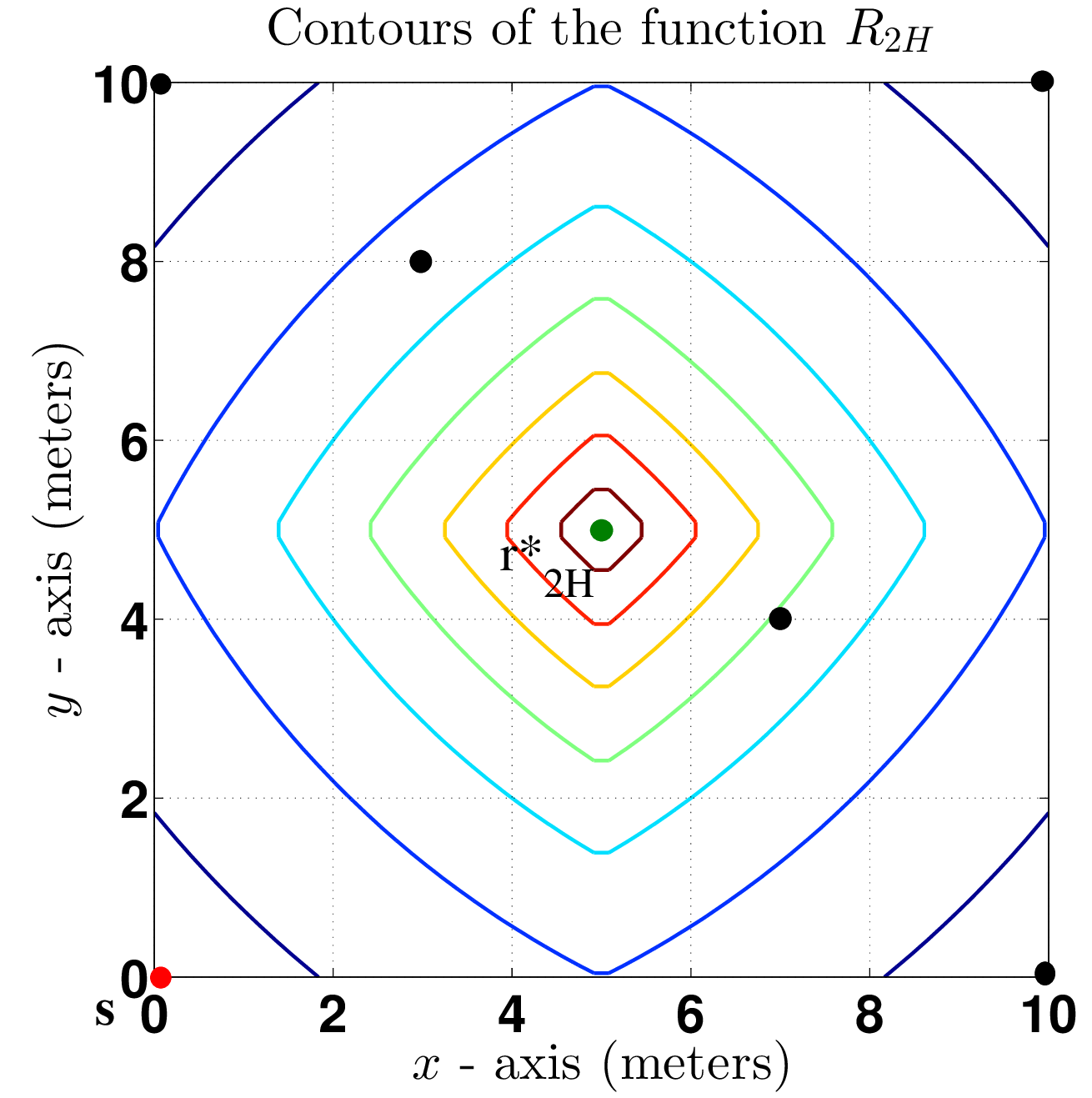}\label{fig:RDFContour}}
\caption{{({\bf a}) $R_{DF}$ for $N=5$; ({\bf b}) $R_{DF}$ contour plot; ({\bf c}) $R_{2H}$ for the same network; ({\bf d}) $R_{2H}$ contour plot.}}
\label{fig6}
\end{figure*}

Consider $N=5$ destinations positioned on a square in the two-dimensional Euclidean
plane with the source node at the origin.
Figure~\ref{fig6}a plots the node positions as circles, and the non-coherent $R_{DF}$ as a function of the relay position.
The best relay position is shown by a circle labeled $\mathbf{r}^*_{\text{DF}}$ and the corresponding rate is
$R_{DF} \approx 0.011 P$ nats/sec.
Figure~\ref{fig6}c plots the low SNR two-hop rate
\begin{align}
R_{2H} \rightarrow \min_{1 \le j \le 5} \frac{1}{2} \min \left(\frac{\xi_{s,r} P_s}{\|\mathbf{s} - \mathbf{r}\|^{\alpha}},
\frac{\xi_{r,j} P_r}{\|\mathbf{r} - \mathbf{j}\|^{\alpha}} \right)
\end{align}
as a function of the relay position. The best relay position is shown by a circle labeled $\mathbf{r}_{2H}^*$ and
the corresponding two-hop rate is $R_{2H} = 0.01 P$ nats/sec. The non-coherent DF gain is thus $\approx$10\%.

Figure~\ref{fig6}b,d  shows contour plots for $R_{DF}$ and $R_{2H}$.
The contours form convex regions, as predicted by Theorem~\ref{thm:LOSDF}.
Again, the relay position maximizing $R_{DF}$ lies closer to the source than the relay position
maximizing $R_{2H}$.

\section{Conclusions}\label{sec:conclusions}

Various quasi-concavity results were established for AWGN MRCs. In particular, the CS rates are quasi-concave
in the relay position for a fixed correlation coefficient (Theorem~\ref{thm:LOSCS}) and the DF rates are
quasi-concave in the relay position (Theorem~\ref{thm:LOSDF}).
\vspace{6pt}

\section*{Acknowledgment}
M. Thakur and G. Kramer were supported by the German Ministry of Education and
Research in the framework of an Alexander von Humboldt Professorship.

\appendix
\subsection{Covariance Matrices and Concavity} \label{sec:cov-con}
%
The covariance matrix of a real-valued random column vector $\mathbf{V}$ is
\begin{align}
\mathbf{Q}_{\mathbf{V}} = \E{ (\mathbf{V}-\E{\mathbf{V}}) (\mathbf{V} - \E{\mathbf{V}} )^T }.
\end{align}
A useful property of covariance matrices is as follows (see~\cite{Cover06}, p.~684).
If $\mathbf{Q}^*_{\mathbf{V}}$ is a principal minor of $\mathbf{Q}_{\mathbf{V}}$, then the
following function is concave in $\mathbf{Q}_{\mathbf{V}}$:
\begin{align}
f(\mathbf{Q}_{\mathbf{V}}) = \log \frac{\det \mathbf{Q}_{\mathbf{V}}}{\det \mathbf{Q}^*_{\mathbf{V}}}.
\end{align}

\subsection{Concave and Quasi-Concave Functions}\label{sec:CA}
\subsubsection{Compositions Preserving Quasi-Concavity}
\label{sec:CA-comp}
The following compositions preserve quasi-concavity.
\begin{Lemma}\label{lemma:qcprops}
Suppose $f$ and $f_{i}$, $1 \le i \le n$, are quasi-concave, then so are the functions
\begin{enumerate}
\item $h=k_1f + k_2$, where $k_1\ge0$ and $k_2\in\mathbf{R}$; \label{Part-lin}
\item $h=\displaystyle\min_{1 \le i \le n} f_{i}$; \label{Part-min}
\item $h=g \circ f$ where $f$ is quasi-concave and $g$ is non-decreasing; \label{Part-comp}
\item $h(\mathbf{a})= \sup_{\mathbf{b}\in\set{B}} f(\mathbf{a},\mathbf{b})$ where $\set{B}$ is a convex set; \label{Part-sup}
\item $h(\mathbf{a},\mathbf{b})=f(g(\mathbf{a}),\mathbf{b})$ where $g$ is convex and
$f(\tilde{\mathbf{a}},\mathbf{b})$ is non-increasing in $\tilde{\mathbf{a}}$ for fixed $\mathbf{b}$. \label{Part-con}
\end{enumerate}
\end{Lemma}
\begin{proof}
Properties 1)--4) are standard (see~\cite{Boyd-book2004}, {Section} 3.4). For property 5), observe that
\begin{align}
& h(\lambda \mathbf{a}_1 + \bar{\lambda}  \mathbf{a}_2, \lambda \mathbf{b}_1 + \bar{\lambda}  \mathbf{b}_2) \nonumber \\
& = f(g(\lambda \mathbf{a}_1 + \bar{\lambda} \mathbf{a}_2), \lambda \mathbf{b}_1 + \bar{\lambda}  \mathbf{b}_2) \nonumber \\
& \overset{(a)}{\ge} f(\lambda g(\mathbf{a}_1) + \bar{\lambda}  g(\mathbf{a}_2), \lambda \mathbf{b}_1 + \bar{\lambda}  \mathbf{b}_2) \nonumber \\
& \overset{(b)}{\ge} \min\left( f(g(\mathbf{a}_1),\mathbf{b}_1), f(g(\mathbf{a}_2), \mathbf{b}_2) \right)
\end{align}
where $(a)$ follows because
$g(\lambda \mathbf{a}_1 + \bar{\lambda}  \mathbf{a}_2) \le \lambda g(\mathbf{a}_1) + \bar{\lambda}  g(\mathbf{a}_2)$
and $f(\tilde{\mathbf{a}},\mathbf{b})$ is non-increasing in $\tilde{\mathbf{a}}$. Step $(b)$ follows because
$f$ is quasi-concave.
\end{proof}

\subsubsection{Examples of Quasi-Concave Functions}
\label{sec:CA-examples}
We establish quasi-concavity for several useful functions.
\begin{Lemma} \label{lemma:qcfunc1}
The following functions are quasi-concave for $\mathbf{x}=(a \; b)$ with non-negative entries.
\begin{enumerate}
\item $f(\mathbf{x}) = ab$
\item $f(\mathbf{x})= \frac{ab}{a+b+k}$ for a positive constant $k$
\item $f(\mathbf{x})=k_1/a+2\sqrt{k_2 b / a}$ for positive constants $k_1,k_2$
\item $f(\mathbf{x})=-(1-b)(k_1+k_2 / a)$ for positive constants $k_1,k_2$, and $b\le1$
\end{enumerate}
Furthermore, the following function is quasi-concave for $\mathbf{x}=(a \; b \; c)$ with non-negative entries.
\begin{enumerate} \setcounter{enumi}{4}
\item $f(\mathbf{x}) = a + b + 2\sqrt{abc}$
\end{enumerate}
\end{Lemma}
\begin{proof}
{We consider positive $\mathbf{x}$, and we} use bordered Hessians $\mathbf{B}_f(\mathbf{x})$
and the derivatives $D_k$ of their $k$th leading principal minors, $k=2,3,\ldots,n$.
The results extend to non-negative $\mathbf{x}$ by using continuity at zero values,
except for the third and fourth functions where $a=0$ makes the functions undefined.
\begin{enumerate}
\item We have $D_2<0$ and $D_3>0$ for
\begin{align*}
\mathbf{B}_f(\mathbf{x}) =
& \begin{pmatrix}
0	& b		& a	\\
b	& 0		& 1	\\
a	& 1		& 0	\\
\end{pmatrix}.
\end{align*}
\item We have $D_2<0$ and $D_3>0$ for
\begin{align*}
\mathbf{B}_f(\mathbf{x}) =
& \begin{pmatrix}
0					& \frac{b(b+k)}{(a+b+k)^2}			& \frac{a(a+k)}{(a+b+k)^2} \\
\frac{b(b+k)}{(a+b+k)^2}	& \frac{-2b(b+k)}{(a+b+k)^3}		& \frac{2ab+(a+b+k)k}{(a+b+k)^3} \\
\frac{a(a+k)}{(a+b+k)^2}	& \frac{2ab+(a+b+k)k}{(a+b+k)^3}	& \frac{-2a(a+k)}{(a+b+k)^3}
\end{pmatrix}.
\end{align*}
\item We have $D_2<0$ and $D_3>0$ for
\begin{align*}
\mathbf{B}_f(\mathbf{x}) =
& \begin{pmatrix}
0						& -\frac{k_1 +\sqrt{k_2 a b}}{a^2}		& \sqrt{\frac{k_2}{ab}} \\
-\frac{k_1 +\sqrt{k_2 a b}}{a^2}	& \frac{4k_1 + 3 \sqrt{k_2 a b}}{2a^3}	& -\frac{\sqrt{k_2}}{2 a^{3/2} \sqrt{b}} \\
\sqrt{\frac{k_2}{ab}}			& -\frac{\sqrt{k_2}}{2 a^{3/2} \sqrt{b}}		& -\frac{\sqrt{k_2}}{2 b^{3/2} \sqrt{a} }
\end{pmatrix}.
\end{align*}
\item If $b \le 1$, we have $D_2<0$ and $D_3>0$ for
\begin{align*}
\mathbf{B}_f(\mathbf{x}) =
& \begin{pmatrix}
0				& \frac{(1-b)k_2}{a^2}	& k_1+\frac{k_2}{a} \\
\frac{(1-b)k_2}{a^2}	& -\frac{2(1-b)k_2}{a^3}	& -\frac{k_2}{a^2} \\
k_1+\frac{k_2}{a}	& -\frac{k_2}{a^2}		& 0
\end{pmatrix}.
\end{align*}
\item We have $D_2<0$, $D_3>0$ and $D_4<0$ for
\begin{align}
\mathbf{B}_f(\mathbf{x}) =
\begin{pmatrix}
0				& 1+\sqrt{\frac{bc}{a}}		& 1+\sqrt{\frac{ac}{b}}		& \sqrt{\frac{ab}{c}} \\
1+\sqrt{\frac{bc}{a}}	& - \frac{\sqrt{bc}}{2 a^{3/2}} 	& \frac{1}{2} \sqrt{\frac{c}{ab}}	& \frac{1}{2} \sqrt{\frac{b}{ac}} \\
1+\sqrt{\frac{ac}{b}}	& \frac{1}{2} \sqrt{\frac{c}{ab}}	& - \frac{\sqrt{ac}}{2 b^{3/2}}	& \frac{1}{2} \sqrt{\frac{a}{bc}} \\
\sqrt{\frac{ab}{c}} 	& \frac{1}{2} \sqrt{\frac{b}{ac}}	& \frac{1}{2} \sqrt{\frac{a}{bc}}	& - \frac{\sqrt{ab}}{2 c^{3/2}}
\end{pmatrix}.
\end{align}
\end{enumerate}
\end{proof}


\end{document}